\documentclass[aps,preprintnumbers,amsmath,amssymb,amsfonts,tightenlines]{revtex4}

\usepackage{bm,slashed,feynmp,graphicx}

\newcommand{\psib}{\bar{\psi}}
\newcommand{\psit}{\widetilde{\psi}}

\newcommand{\dmb}{\partial_{\mu}}
\newcommand{\dmt}{\partial^{\mu}}
\newcommand{\abp}{a_{\bm{p}}}
\newcommand{\bbp}{b_{\bm{p}}}
\newcommand{\id}{\bm{1}_2}
\newcommand{\Gammap}{\Gamma_{p}}
\newcommand{\uhat}{\widetilde{u}}

\newcommand{\cl}{\textrm{cl}}
\newcommand{\dsl}{\slashed{\partial}}

\begin{document}

    \bibliographystyle{apsrev}

    \title{Lorentz Symmetric Quantum Field Theory for Symplectic Fermions}
      	\author{Dean J. Robinson}
                \email[email:]{djr233@cornell.edu}          
   	 \author{Eliot Kapit}
    	\author{Andr\'e LeClair}
                
    \affiliation{Newman Laboratory, Cornell University, Ithaca, N.Y.}
    \date{\today}
     
    \begin{abstract}
      A free quantum field theory with Lorentz symmetry is derived for spin-half symplectic fermions in $2+1$ dimensions. In particular, we show that fermionic spin-half fields may be canonically quantized in a free theory with a Klein-Gordon Lagrangian.  This theory is shown to have all the required properties of a consistent free quantum field theory, namely causality, unitarity, adherence to the spin-statistics theorem, $\mathcal{C}\mathcal{P}\mathcal{T}$ symmetry, and the Hermiticity and positive definiteness of the Hamiltonian. The global symmetry of the free theory is Sp(4) $\simeq$ SO(5). Possible interacting theories of both the pseudo-Hermitian and Hermitian variety are then examined briefly.
    \end{abstract}
\maketitle

\section{Introduction}

Recently a quantum field theory for the so-called symplectic fermions in $2+1$ dimensions was presented \cite{LeClairNeubert:2007sl,KapitLeClair:2009sf}, in which the kinetic terms were second order in derivatives so that the free theory was of Klein-Gordon form rather than the usual Dirac Lagrangian. This theory was developed within the context of high-temperature superconductivity, the main attractive features being that it permitted a renormalizable four-fermion interaction and had a spontaneously broken Sp(4) $\simeq$ SO(5) internal symmetry. This produced both antiferromagnetic and superconducting order parameters lying in the vector representation of a broken SO(5) symmetry as proposed by Zhang \cite{Zhang:1997ut}. However, this theory was manifestly non-relativistic in the sense that the fermion fields were defined as Lorentz scalars and spin identified as a flavor.

In this paper, we derive the corresponding $2+1$ dimensional, Poincar\'e invariant, quantum field theory for symplectic fermions, which properly transform under the fundamental Dirac representation of the Lorentz group. In doing so, we show that fermionic spin-half fields may be canonically quantized in a free theory with a \emph{Klein-Gordon} Lagrangian, contrary to the common belief that this is only possible for a Lagrangian of the Dirac type. That this could be possible without violation of the spin-statistics theorem was considered long ago by Pauli \cite{Pauli:1940ss}, and we show in this paper, by finding an explicit form for the fields, that this free theory indeed has all the required properties of a consistent free quantum field theory. In particular, we show: causality is preserved; the Fock space Hamiltonian and physical momentum take the usual form, the former being both Hermitian and positive definite; the spin-statistics theorem is satisfied; and the theory has separate $\mathcal{C}$, $\mathcal{P}$, and $\mathcal{T}$ symmetries. Unlike in the $2+1$ Dirac theory, the latter property holds for non-zero mass terms. As such, it appears that there exists a class of fermionic Lagrangians with Klein-Gordon kinetic terms that are yet to be explored in the Literature. Although specialized to $2+1$ dimensions in this paper, our derivation for the free theory may also be generalized to higher dimensions.

The paper is structured as follows. In Sec. \ref{sec:TFT} we present the canonically quantized free theory, with details of the novel spinor structure provided in an appendix. Symmetries of the free theory and conserved charges are discussed, together with pseudo-Hermitian properties of the theory. In Sec. \ref{sec:AIT} we then present an example of a pseudo-Hermitan and a Hermitian renormalizable interacting theory involving symplectic fermions. We examine the global and discrete symmetries of the latter theory and show it has separate parity and time-reversal violation. Its Feynman rules are briefly discussed, and the example of electron-electron scattering is briefly considered.

\section{The free theory}
\label{sec:TFT}
\subsection{Lagrangian, fields and spinors}
\label{sec:LFS}

Let $\Lambda$ denote a transformation under the vector representation of the $2+1$ dimensional Lorentz group, SO(2,1), and let $\Lambda_{1/2}$ denote the corresponding element of its fundamental Dirac representation. In order to construct a Lorentz invariant quantum field theory for fermions in this representation, we consider a fermionic field $\psi$ and an adjoint field $\psit$ with Lorentz transformation
\begin{equation}
	\label{eqn:TFF}
	\psi(x) \mapsto \Lambda_{\frac{1}{2}}\psi(\Lambda^{-1}x)~,~~\mbox{and}~ \psit(x) \mapsto \psit(\Lambda^{-1}x)\Lambda_{\frac{1}{2}}^{-1}~.
\end{equation}
Since $\Lambda_{1/2}$ does not act on the Lorentz vector indices, the Klein-Gordon Lagrangian
\begin{equation}
	\label{eqn:LFT}
	\mathcal{L}_{0} = \dmb\psit\dmt\psi - m^2\psit\psi~,
\end{equation}
is manifestly Lorentz invariant and is therefore a candidate Lagrangian for a relativistic free theory of fermions. The corresponding action is of course translation invariant, so such a theory has Poincar\'e symmetry. 

The equation of motion arising from $\mathcal{L}_0$ is a Klein-Gordon equation of the usual form
\begin{equation}
	\label{eqn:FTEM}
	(\partial^2 + m^2)\psi = 0~,
\end{equation}	
and the canonical momentum fields corresponding to $\psi$ and $\psit$ are respectively $\partial_0\psit$ and $\partial_0\psi$, so that the canonical equal-time anticommutation relations for a free quantum theory described by $\mathcal{L}_0$ must be 
\begin{equation}
	\label{eqn:ETARF}
	\big\{\psi_a(x^0,\bm{x}),\partial_0\psit_b(x^0,\bm{y})\big\}  = -\big\{\psit_a(x^0,\bm{x}),\partial_0\psi_b(x^0,\bm{y})\big\} = i\delta^{(2)}(\bm{x} - \bm{y})\delta_{ab}~,
\end{equation}
with all other equal-time anticommutators zero. In order to canonically quantize this theory, we must find fields that satisfy both the equation of motion and the anticommutation relations (\ref{eqn:ETARF}). A correct choice for the fermion field and its adjoint is
	\begin{align}
	\psi(x) & = \int\!\!\!\frac{d^2\bm{p}}{(2\pi)^2}\frac{1}{\sqrt{2E_{\bm{p}}}}\sum_{s=0,1}u^s(p)\Big(\abp^s e^{-ip\cdot x} + \bbp^{s\dagger}e^{ip\cdot x}\Big)~,\notag\\
	\psit(x) & = \int\!\!\!\frac{d^2\bm{p}}{(2\pi)^2} \frac{1}{\sqrt{2E_{\bm{p}}}}\sum_{s=0,1}\uhat^s(p)\Big(\abp^{s\dagger} e^{ip\cdot x} - \bbp^{s} e^{-ip\cdot x}\Big)~, \label{eqn:CQ}
	\end{align}
where the usual dispersion relation $p^2 = m^2$ holds. In Eqs. (\ref{eqn:CQ}), we have $E_{\bm{p}} = \sqrt{\bm{p}^2 + m^2}$ and the operators $\abp^s$ and $\bbp^s$ are Fock space annihilation operators satisfying the usual anticommutation relations
\begin{equation}
	\label{eqn:ETAR}
	\big\{\abp^r,a_{\bm{q}}^{s\dagger}\big\} =\big\{\bbp^r,b_{\bm{q}}^{s\dagger}\big\} = (2\pi)^2\delta^{rs}\delta^{(2)}(\bm{p} - \bm{q})~,
\end{equation}
with all other anticommutators zero. As shown in Appendix \ref{app:SS}, the basis spinors $u^s(p)$ may be chosen such that they satisfy a Dirac equation 
\begin{equation}
	\label{eqn:SDE}
	\big(\slashed{p}/m\big)u^s(p) =  (-1)^{s}u^s(p)~,
\end{equation}
where $\slashed{p}$ is written in the Feynman slash notation, and the adjoint spinor $\uhat^s(p)$ may then be defined by
\begin{equation}
	\label{eqn:SR1}
	\uhat^s(p) \equiv (-1)^s\bar{u}^s(p)~,
\end{equation}
where $\bar{u}^s(p)$ is the usual Dirac adjoint spinor (henceforth an overline always denotes the Dirac adjoint). It is important to note that Eqs. (\ref{eqn:SDE}) and (\ref{eqn:SR1}) hold only for a special choice of the basis spinors (see App. \ref{app:SS}): here and henceforth $u^s(p)$ denotes the choice of basis as defined in Eqs. (\ref{eqn:DGZE}) and (\ref{eqn:DTL}). 

One may show (see App. \ref{app:SS}) that $\uhat^s(p)$ is a well-defined adjoint in the sense that the inner product $\uhat^s(p) u^s(p)$ is positive definite and Lorentz invariant. In particular, the spinor and adjoint spinor Lorentz transform as
\begin{equation}
	\label{eqn:SLT}
	u^s(\Lambda p) = \Lambda_{\frac{1}{2}}u^s(p)~,~~\mbox{and}~\uhat^s(\Lambda p) = \uhat^s(p)\Lambda_{\frac{1}{2}}^{-1}~,
\end{equation}
the inner product is in fact simply
\begin{equation}
	\label{eqn:SNR}
	\uhat^r(p)u^s(p) = \delta^{rs}~,
\end{equation}
and one also finds the trivial completeness relations
\begin{equation}
	\label{eqn:SCR}
	\sum_{s= 0,1}u_a^s(p)\uhat_b^s(p) = \delta_{ab}~.
\end{equation}
Using the completeness relations (\ref{eqn:SCR}), we may calculate easily the free-field Feynman prescription propagator, with the result that
\begin{equation}
	\label{eqn:FTFP}
	\langle0|T\psi(x)\psit(y)|0\rangle = \id\int\frac{d^3p}{(2\pi)^3}\frac{i}{p^2-m^2 + i\varepsilon}e^{-ip\cdot(x-y)}~,
\end{equation}
where $T$ denotes the usual fermionic time ordering operator, $\id$ is the 2$\times$2 identity, and the integral over $p^0$ has Feynman boundary conditions. Similarly, the anticommutator $\{\psi_a(x),\psit_b(y)\} = 0$ outside the lightcone, so causality is preserved.
 
It is perhaps instructive to rewrite Eqs. (\ref{eqn:CQ}) in the familiar notation of the $2+1$ Dirac theory, which can be achieved by noting the correspondence $u(p) = u^0(p)$ and $v(p) = u^1(p)$ implied by the Dirac equation (\ref{eqn:SDE}) and then applying Eq. (\ref{eqn:SR1}). In this notation, the adjoint field becomes
\begin{equation}
	\label{eqn:AFDN}
	\psit(x)  = \int\!\!\!\frac{d^2\bm{p}}{(2\pi)^2} \frac{1}{\sqrt{2E_{\bm{p}}}} \Big( \big[\bar{u}(p)\abp^{0\dagger} - \bar{v}(p)\abp^{1\dagger}\big]e^{ip\cdot x} + \big[\bar{v}(p)\bbp^{1}-\bar{u}(p)\bbp^{0}\big]e^{-ip\cdot x}\Big)~.
\end{equation}
It can be seen from this that the adjoint field consists of a Dirac adjoint field with two extra degrees of freedom, whose terms have minus signs and the positive (negative) energy term has the negative (positive) energy Dirac spinor.  It is also clear in this expanded form that the minus sign in Eq. (\ref{eqn:CQ}) ensures that the sign in front of the $\bar{v}(p)\bbp^1$ term is as would be expected in the Dirac theory. One can carry out the analysis of the free theory using this formulation together with the completeness relations in Eq. (\ref{eqn:SR2}). However this notation is obviously less compact.

\subsection{Hamiltonian and physical momentum}
The canonical Hamiltonian density $\mathcal{H}_0 = \partial_0\psit\partial_0\psi + \nabla\psit\cdot\nabla\psi + m^2\psit\psi$. Applying the normalization relations (\ref{eqn:SNR}) and the dispersion relation $p^2 = m^2$, one finds that the free theory Hamiltonian is simply
\begin{equation}
	\label{eqn:HFT}
	H_0 = \int d^2 \bm{x}\mathcal{H}_0 = \int\frac{d^2\bm{p}}{(2\pi)^2}\sum_{s=0,1}E_{\bm{p}}\Big(\abp^{s\dagger}\abp^s + \bbp^{s\dagger}\bbp^s\Big)~,
\end{equation}
up to an infinite additive constant. Here the cross terms $\propto \abp^{s\dagger}\bbp^{r\dagger}$ or $\bbp^s\abp^r$ cancel via the dispersion relation, just as in the scalar Klein-Gordon theory. This Hamiltonian is clearly Hermitian and has a positive definite energy spectrum. Further, from Eqs. (\ref{eqn:ETAR}) it is clear that all states have positive norm.

The stress-energy tensor generated by the translation invariance of the free theory is 
\begin{equation}
\label{eqn:SET}
	T_{\mu\nu} = \partial_\mu\psit\partial_\nu \psi + \partial_\nu \psit \partial_\mu \psi - \eta_{\mu\nu}\mathcal{L}_0
\end{equation}
where $\eta_{\mu\nu}$ is the Minkowski metric. As usual, the physical momentum $P_\mu \equiv \int d^2\bm{x} T_{\mu 0}$ and clearly $\mathcal{H}_0 = T_{00}$ so that $P_0 = H_0$. Similarly, the spatial components
\begin{align}
	P_i 
	& = \int d^2\bm{x}\big(\partial_i \psit\partial_0 \psi + \partial_0 \psit \partial_i \psi\big)\notag\\
	& = \int\!\!\frac{d^2\bm{p}}{(2\pi)^2} \sum_{s=0,1} p_i \Big(\abp^{s\dagger}\abp^s + \bbp^{s\dagger}\bbp^s\Big)~
\end{align}
up to an infinite additive constant. Hence the physical momentum
\begin{equation}
	P_\mu = \int\!\!\frac{d^2\bm{p}}{(2\pi)^2} \sum_{s=0,1} p_\mu \Big(\abp^{s\dagger}\abp^s + \bbp^{s\dagger}\bbp^s\Big)~.
\end{equation}
Note that clearly $[P_\mu,P_\nu] = 0$ and that $P_\mu$ is the generator of translations.

As a check of the Lorentz invariance, just as for the Dirac theory the Lorentz transformations (\ref{eqn:TFF}) together with Eqs. (\ref{eqn:SLT}) and the Lorentz invariant measure $d^2\bm{p}/[(2\pi)^2 2E_{\bm{p}}]$ imply that 
\begin{equation}
	U\abp^sU^{-1} = \sqrt{\frac{E_{\Lambda \bm{p}}}{E_{\bm{p}}}}a^s_{\Lambda \bm{p}}~,
\end{equation}
and similary for other operators, where $U$ implements the Lorentz transformation on the Fock space. It follows that
\begin{equation}
	UP_{\mu}U^{-1} = \int\!\!\frac{d^2(\Lambda\bm{p})}{(2\pi)^2} \sum_{s=0,1} p_\mu \Big(a_{\Lambda \bm{p}}^{s\dagger}a_{\Lambda \bm{p}}^s + b_{\Lambda \bm{p}}^{s\dagger}b_{\Lambda \bm{p}}^s\Big) = \big(\Lambda^{-1}\big)^{~\nu}_\mu P_\nu~,
\end{equation}
as expected for a Lorentz invariant theory. 

\subsection{Generalized adjoint field and Hermiticity}
Combining Eqs. (\ref{eqn:SDE}) and (\ref{eqn:SR1}) (or more precisely Eqs. (\ref{eqn:DGAS}) and (\ref{eqn:CGM})) we have $\uhat^s(p) = [(\slashed{p}/m)u^s(p)]^\dagger\gamma^0$. It is worth noting that upon inserting this expression into Eqs. (\ref{eqn:CQ}) it follows immediately that we may write
\begin{equation}
	\label{eqn:DAF}
	\psit= \overline{[i\dsl/m]\psi}~.
\end{equation}
In the case that $\psi$ belongs to the subspace of solutions of Eq. (\ref{eqn:FTEM}) which satisfy the Dirac equation, we obtain from Eq. (\ref{eqn:DAF}) simply $\psit = \psib$, exactly as expected: As is well-known, solutions of the Dirac equation must satisfy the Klein-Gordon equation. Equation (\ref{eqn:DAF}) implies that $\psit$ is not just an \emph{ad hoc} field satisfying Eqs. (\ref{eqn:FTEM}) and (\ref{eqn:ETARF}), but rather an extension of the Dirac adjoint $\psib$ to the full solution space of the adjoint equation of motion (see App. \ref{app:SS}). It is important to note that the differential operator and mass in Eq. (\ref{eqn:DAF}) do not play a canonical r\^ole in the Lagrangian itself, but are instead just an effective means of bookkeeping the various minus signs in Eq. (\ref{eqn:AFDN}): $\psit$ remains the canonical adjoint field to $\psi$.

Let us now extend Eq. (\ref{eqn:DAF}) and henceforth use it to define a generalized adjoint field $\psit$ for any arbitrary field $\psi$. In the case that $\psi$ satisfies the free theory equation of motion - i.e. it is a free field - $\psit$ then reduces to the adjoint free field (\ref{eqn:CQ}) presented above. Further, under this new definition, $\psit$ manifestly transforms under (global) Lorentz transformations according Eq. (\ref{eqn:TFF}), because $\dsl$ is a tensor of the spin indices (see App. \ref{app:SS} and Eq. (\ref{eqn:ASTL})). As such, in terms of the free theory and Lorentz invariance the adjoint field $\psit$ is well-defined by Eq. (\ref{eqn:DAF}).

Importantly, the field product $\psit(x)\psi(x)$ is now Hermitian by construction in the same sense that the kinetic term of the Dirac Lagrangian is Hermitian: they are both Hermitian up to a total derivative. Equivalently, via integration by parts
\begin{align}
	\int d^3x \big[\psit(x)\psi(x)\big]^\dagger
	& = \int d^3x\big[\psi^\dagger\gamma^0(i\dsl/m)\psi\big]\notag\\
	& = -\int d^3x \big[i(\dmb/m)\psi^\dagger\gamma^{\mu\dagger}\gamma^0\psi\big]\notag\\
	& = \int d^3x \big[\psit(x)\psi(x)\big]~,
\end{align}
where we have used $\gamma^{\mu\dagger} = \gamma^0\gamma^\mu\gamma^0$. It immediately follows from this result that for any field $\psi$ and an adjoint defined by Eq. (\ref{eqn:DAF}), the kinetic and mass terms in the action $\int d^3x\mathcal{L}_0$ or the Hamiltonian $H_0$ are independently Hermitian, as is any term of the form $\int d^3x \mathcal{O}\psit\mathcal{O}\psi$ for an operator $\mathcal{O}$ that commutes with $\partial_\mu$. In Eq. (\ref{eqn:HFT}) we explicitly verified that $H_0$ was Hermitian and positive definite when expanded in terms of the free field creation and annihilation operators. It is now clear that Hermiticity of the free theory action and $H_0$ may be guaranteed by construction, whether or not the free theory equation of motion is satisfied. This implies that we may in principle construct interacting theories with real spectra.

\subsection{Global continuous symmetries, fermion number and spin}
Let us consider the global continuous symmetries of the free theory $\mathcal{L}_0$. First, it is clear that the free theory has a U(1) symmetry $\psi \to e^{i\alpha}\psi$. Applying the Fermi statistics $\{\psi_a,\psit_b\} = 0$, the conserved charge for this symmetry is
\begin{equation}
	Q  = i\int d^2\bm{x}\big(\partial^0\psit\psi - \psit\partial^0\psi\big)~.
\end{equation}
Expanding in terms of creation and annihilation operators via Eqs. (\ref{eqn:CQ}), cross terms again cancel yielding
\begin{equation}
	Q = \int\frac{d^2\bm{p}}{(2\pi)^2}\sum_{s=0,1}\Big(\abp^{s\dagger}\abp^s - \bbp^{s\dagger}\bbp^s\Big)~,
\end{equation}
up to an infinite additive constant. $Q$ is the usual fermion number, such that $Q = +1$ for particles (created by $\abp^{s\dagger}$) and $Q = -1$ for antiparticles (created by $\bbp^{s\dagger}$).

The free theory also has a global Lorentz symmetry SO(2,1) and Poincar\'e symmetry by construction. In particular, it remains for us to check explicitly that the single (anti)particle states $\abp^{s\dagger}|0\rangle$ ($\bbp^{s\dagger}|0\rangle$) are spin-half states. In $2+1$ dimensions the angular momentum operator, $J$, is the conserved charge corresponding to the symmetry generated by $S^{12} = \gamma^0/2$. In the rest frame, the infinitesmal symmetry is $\delta \psi = -(i/2)\theta \gamma^0\psi$ and $\delta \psit = +(i/2)\theta \psit\gamma^0$ so that
\begin{equation}
	J = -\frac{i}{2}\int d^2\bm{x} \big(\psit\gamma^0\partial_0\psi - \partial_0\psit\gamma^0\psi\big)~.
\end{equation}
Inserting the quantized fields (\ref{eqn:CQ}) we then find
\begin{equation}
	\label{eqn:FEAO}
	J = \frac{1}{2}\int\frac{d^2\bm{p}}{(2\pi)^2}\sum_{r,s = 0,1}\uhat^{r}(p)\gamma^0u^s(p)\big(\abp^{r\dagger}\abp^s + \bbp^r\bbp^{s\dagger}\big)~.
\end{equation}
Applying the spinor relation in Eq. (\ref{eqn:SR4}) we may verify $J|0\rangle = 0$, and it follows further from Eq. (\ref{eqn:FEAO}) that $a_{\bm{0}}^{0\dagger}|0\rangle$ ($a_{\bm{0}}^{1\dagger}|0\rangle$) is a spin $+1/2$ ($-1/2$) particle state and $b_{\bm{0}}^{0\dagger}|0\rangle$ ($b_{\bm{0}}^{1\dagger}|0\rangle$) is an  antiparticle state of opposite spin $-1/2$ ($+1/2$). Hence the spin statistics theorem is satisfied, as expected. Note that the $a_{\bm{0}}^{0\dagger}$ and $b_{\bm{0}}^{1\dagger}$ operators belong to the terms which form the subspace of solutions satisfying the Dirac equation, as can be seen in Eq. (\ref{eqn:AFDN}). These operators both form spin $+1/2$ or `spin up' states just as in the $2+1$ Dirac theory, so we may also think of $\psi$ and $\psit$ to be extensions of the Dirac fermions to include `spin down' states, though the symplectic fermions obviously have different mass dimensions.

The maximal global symmetry of the free theory is Sp(4), which can be seen as follows. Let
\begin{equation}
	\label{eqn:SSFT}
	\Psi = \begin{pmatrix}  \psit^T \\ \psi~ \end{pmatrix}~,
\end{equation}
where the superscript $T$ denotes here the transpose. Applying the Fermi statistics  we may then rewrite the free Lagrangian as
\begin{equation}
	\mathcal{L}_{\textrm{free}} = \frac{1}{2}\Big(\!\dmb\Psi^TJ\dmt\Psi - m^2\Psi^TJ\Psi\!\Big),~ J = \begin{pmatrix} 0 & \id \\ -\id & 0 \end{pmatrix}~,
\end{equation}	
so that clearly $\mathcal{L}_\textrm{free}$ has an Sp(4, $\mathbb{C}$) internal symmetry. Note that this argument can be extended to $N$ component spinors, with corresponding symmetry Sp(2$N$, $\mathbb{C}$). Further, it is straightforward to check that the abovementioned global symmetries U(1)$\times$SO(2,1) $\subset$ Sp(4,$\mathbb{C}$) by writing them as linear transformations of $\Psi$.

\subsection{Discrete symmetries}
In $2+1$ dimensions there are two parity operators, namely $\mathcal{P}_1: (x^1,x^2) \mapsto (-x^1,x^2)$ and $\mathcal{P}_2: (x^1,x^2) \mapsto (x^1,-x^2)$. Note that $\mathcal{R}: \bm{x} \to -\bm{x}$ has determinant $+1$ and is equivalent to a rotation, so that $\mathcal{P}_1 \propto \mathcal{R}\mathcal{P}_2$. Since angular momentum in $2+1$ is the scalar $x^1p^2 - p^1x^2$, the parity transformations $\mathcal{P}_{1,2}$ must both change the sense of rotation and hence we expect parity to flip spins. Choosing the $\gamma^\mu$ matrices to be explicitly $\gamma^0 = \sigma_z$, $\gamma^1 = i\sigma_x$ and $\gamma^2 = i\sigma_y$, where $\sigma_{x,y,z}$ are the usual $2\times 2$ Pauli matrices, we find the parity operators act on free fields as
\begin{align}
	\mathcal{P}_1\psi\mathcal{P}_1^\dagger & = \gamma^1\psi, & \mathcal{P}_1\psit\mathcal{P}_1^\dagger  & = -\psit\gamma^1\notag\\
	\mathcal{P}_1\abp^s\mathcal{P}_1^\dagger & = ia_{\bm{p}_1}^{1-s}, & \mathcal{P}_1\bbp^s\mathcal{P}_1^\dagger & = -ib_{\bm{p}_1}^{1-s}\notag\\
	\mathcal{P}_2\psi\mathcal{P}_2^\dagger & = i\gamma^2\psi, & \mathcal{P}_2\psit\mathcal{P}_2^\dagger  & = i\psit\gamma^2\notag\\
	\mathcal{P}_2\abp\mathcal{P}_2^\dagger & = i(-1)^sa_{\bm{p}_2}^{1-s}, & \mathcal{P}_2\bbp\mathcal{P}_2^\dagger & = -i(-1)^sb_{\bm{p}_2}^{1-s}~,\label{eqn:POA}
\end{align}
where $\bm{p}_1 = (-p^1,p^2)$ and $\bm{p}_2 = (p^1,-p^2)$. Further, one finds that the anti-linear time-reversal operator $\mathcal{T}$  and charge conjugation operator $\mathcal{C}$ act as
\begin{align}
	\mathcal{T}\psi\mathcal{T}^\dagger & = i\gamma^2\psi, & \mathcal{T}\psit\mathcal{T}^\dagger  & = i\psit\gamma^2\notag\\
	\mathcal{T}\abp^s\mathcal{T}^\dagger & = i(-1)^sa^{1-s}_{-\bm{p}}, & \mathcal{T}\bbp^s\mathcal{T}^\dagger & = -i(-1)^sb^{1-s}_{-\bm{p}}\notag\\
	\mathcal{C}\psi\mathcal{C}^\dagger & = \gamma^1\psi^*, & \mathcal{C}\psit\mathcal{C}^\dagger & = \psit^*\gamma^1\notag\\
	\mathcal{C}\abp^s\mathcal{C}^\dagger & = i\bbp^{1-s}, & \mathcal{C}\bbp^s\mathcal{C}^\dagger & = -i\abp^{1-s}~.\label{eqn:TCOA}
\end{align}
To verify that $\mathcal{P}_{1,2}$, $\mathcal{T}$, and $\mathcal{C}$ are well-defined operators on the Fock space, note that we may write
\begin{align}
	\mathcal{P}_1 &=  \exp\bigg[ -\frac{i\pi}{2}\int\!\!\frac{d^2\bm{p}}{(2\pi)^2}\sum_{s=0,1}\Big(a^{(1-s)\dagger}_{\bm{p}_1}\abp^s - b^{(1-s)\dagger}_{\bm{p}_1}\bbp^s\Big)\bigg]~,\notag\\
	\mathcal{P}_2 &=  \exp\bigg[ -\frac{i\pi}{2}\int\!\!\frac{d^2\bm{p}}{(2\pi)^2}\sum_{s=0,1}\Big(a^{(1-s)\dagger}_{\bm{p}_2}\abp^s - b^{(1-s)\dagger}_{\bm{p}_2}\bbp^s\Big)\bigg] \notag\\
	& \quad \times \exp\bigg[ i\pi\int\!\!\frac{d^2\bm{p}}{(2\pi)^2}\sum_{s=0,1}s\Big(\abp^{s\dagger}\abp^s + \bbp^{s\dagger}\bbp^s\Big)\bigg]~,\notag\\
	\mathcal{T} &=  \exp\bigg[ -\frac{i\pi}{2}\int\!\!\frac{d^2\bm{p}}{(2\pi)^2}\sum_{s=0,1}\Big(a^{(1-s)\dagger}_{-\bm{p}}\abp^s - b^{(1-s)\dagger}_{-\bm{p}}\bbp^s\Big)\bigg] \notag\\
	& \quad \times \exp\bigg[ i\pi\int\!\!\frac{d^2\bm{p}}{(2\pi)^2}\sum_{s=0,1}s\Big(\abp^{s\dagger}\abp^s + \bbp^{s\dagger}\bbp^s\Big)\bigg]~,\notag\\
	\mathcal{C} &=  \exp\bigg[ -\frac{i\pi}{2}\int\!\!\frac{d^2\bm{p}}{(2\pi)^2}\sum_{s=0,1}\Big(b^{(1-s)\dagger}_{\bm{p}}\abp^s + a^{(1-s)\dagger}_{\bm{p}}\bbp^s\Big)\bigg]\notag\\
	& \quad \times \exp\bigg[ i\pi\int\!\!\frac{d^2\bm{p}}{(2\pi)^2}\sum_{s=0,1} \bbp^{s\dagger}\bbp^s\bigg]~,\label{eqn:FSRDS}
\end{align}
from which it is clear that all these operators are unitary. Since $(\gamma^1)^2 = (\gamma^2)^2 = -1$ and $\dmb\dmt$ is an invariant of these operators, then both parity and time-reversal are symmetries of the free theory. This is unlike the case of the $2+1$ Dirac theory, in which a mass term breaks parity and time-reversal invariance (see e.g. Ref. \cite{JackiwTempleton:1981tp}). For the charge conjugation operator, we have via the Fermi statistics
\begin{equation}
	\label{eqn:CAFP}
	\mathcal{C}\psit\psi\mathcal{C}^\dagger = -\psit^*\psi^* = [\psit\psi]^\dagger \not= \psit\psi~.
\end{equation}
However, the Hermiticity of $H_0$ implies that $\mathcal{C}H_0\mathcal{C}^\dagger = H_0$ and so $\mathcal{C}$ is a symmetry of the free theory. One may also see this directly from the action of $\mathcal{C}$ on the Fock operators in Eq. (\ref{eqn:HFT}). Hence the $\mathcal{C}\mathcal{P}\mathcal{T}$ theorem is verified.

\subsection{Pseudo-Hermiticity}
As mentioned above, the product of arbitrary fields $\psit(x)\psi(x)$ is Hermitian only up to a total derivative and hence not Hermitian. This can be seen explicitly for the free fields by noting either Eq. (\ref{eqn:SR3}) or the extra minus sign in Eq. (\ref{eqn:CQ}). However, in the case that $\psi$ and $\psit$ are free fields it can be shown that this product is instead pseudo-Hermitian.

In Appendix \ref{app:SS} we considered an ansatz for the classical adjoint field $\psit_{\cl}$ (\ref{eqn:GSEMFT2}). The usual canonical quantization of $\psi_{\cl}$ and $\psit_{\cl}$ involves simply replacing the Fourier coefficients $\alpha^s_{\bm{p}}$ and $\beta^s_{\bm{p}}$ in Eqs. (\ref{eqn:GSEMFT}) and (\ref{eqn:GSEMFT2}) with Fock space operators to form the adjoint field
\begin{equation}
	\label{eqn:NAF}
	\psit_{\textrm{0}}(x) = \int\!\!\!\frac{d^2\bm{p}}{(2\pi)^2} \frac{1}{\sqrt{2E_{\bm{p}}}}\sum_{s=0,1}\uhat^s(p)\Big(\abp^{s\dagger} e^{ip\cdot x} + \bbp^{s} e^{-ip\cdot x}\Big)~,
\end{equation}
but this adjoint does not satisfy the equal-time anticommutation relations (\ref{eqn:ETARF}). Instead, similarly to the quantization procedure devised for the scalar sympletic fermion theory \cite{LeClairNeubert:2007sl, KapitLeClair:2009sf}, let $C$ be a Hermitian unitary operator which acts upon the Fock space such that
\begin{equation}
	\label{eqn:DCO}
	C\bbp^s C = -\bbp^s~, ~~ C\abp^s C = \abp^s~,~~\mbox{and}~C|0\rangle = |0\rangle~.
\end{equation}
Note that $C$ is well-defined as a Fock space operator, since we may write
\begin{equation}
	\label{eqn:DCOG}
	C = \exp\bigg[ i\pi \int\!\!\frac{d^2\bm{p}}{(2\pi)^2}\sum_{s=0,1}\bbp^{s\dagger}\bbp^s\bigg]~,
\end{equation}
and explicitly $C = C^{-1} = C^\dagger$. The adjoint field defined by
\begin{equation}
	\label{eqn:DPSA}
	\psit(x) \equiv C\psit_{0}(x) C~,
\end{equation}	
is simply the adjoint in Eq. (\ref{eqn:CQ}). Together with the definition of the annihilation operators $\abp^s$ and $\bbp^s$, Eq. (\ref{eqn:DPSA}) prescribes the canonical quantization procedure of the free theory. It is also interesting to note that from Eq. (\ref{eqn:FSRDS}) $C$ appears to be a constituent of the charge conjugation operator $\mathcal{C}$.

Similarly, let $S$ be the Hermitian, unitary, operator $S = S^{-1} = S^{\dagger}$, which acts upon the Fock space and vacuum $|0\rangle$ such that
\begin{equation}
	\label{eqn:DSO}
	S\abp^{s}S = (-1)^s\abp^s~,~~S\bbp^{s}S = (-1)^s\bbp^s~,~~\mbox{and}~S|0\rangle = |0\rangle~.
\end{equation}
With reference to Eq. (\ref{eqn:SR1}) it is clear that $\psit_{0} = S\psib S$ (see App. \ref{app:SS}). Noting that $C$ and $S$ clearly commute, let $\eta \equiv CS$, so that
\begin{equation}
	\label{eqn:DEO}
	\eta\abp^{s}\eta = (-1)^s\abp^s~,~~\eta\bbp^{s}\eta = (-1)^{1-s}\bbp^s~,~~\eta^2 = 1.
\end{equation}
Explicitly as a Fock space operator
\begin{equation}
	\label{eqn:DEOG}
	\eta = \exp\bigg[i\pi\int\!\!\frac{d^2\bm{p}}{(2\pi)^2}\bigg(\abp^{1\dagger}{\abp^1} + \bbp^{0\dagger}\bbp^{0}\bigg)\bigg]~,
\end{equation}
so $\eta$ is Hermitian and unitary. It follows from Eq. (\ref{eqn:DPSA}) that 
\begin{equation}
	\label{eqn:EADA}
	\psit = \eta\psib \eta~,
\end{equation}
and from Eq. (\ref{eqn:DEOG}) it is clear that $\eta$ simply swaps the signs in front of the spin down states $\bbp^0$ and $\abp^1$.

The Hermiticity and unitary of $\eta$ guarantees that
\begin{equation}
	\label{eqn:EHE}
	[\psit\psi]^\dagger = \psib \eta\psi \eta = \eta\psit\psi \eta~.
\end{equation}		
Hence $\psit\psi$ is pseudo-Hermitian with respect to $\eta$ (henceforth $\eta$-pseudo-Hermitian), as claimed.  Further, comparing Eqs. (\ref{eqn:DAF}) and (\ref{eqn:EADA}) and noting again that $\eta$ is Hermitian and unitary, we may deduce the action of $\eta$ on the free field $\psi$
\begin{equation}
	\label{eqn:EAF}
	\eta \psi \eta = (i\slashed{\partial}/m)\psi~,
\end{equation}
from which we see that $\eta^2 = 1$ is equivalent to the free theory equation of motion (\ref{eqn:FTEM}).

\section{Interacting Theories}
\label{sec:AIT}
\subsection{Pseudo-Hermitian interactions}
Having presented the free theory (\ref{eqn:LFT}) above, we now present examples of interacting theories for the symplectic fermions. Hermiticity contraints due to Eq. (\ref{eqn:EHE}) limit the possible number of Hermitian interacting Hamiltonians. For example, since the spinors have two components, the quartic term $(\psit\psi)^2$ is non-zero and has mass dimension two, so it may act as a renormalizable interaction term. However, as in Eq. (\ref{eqn:EHE}), such a product of the interaction picture fields is $\eta$-pseudo-Hermitian, rendering the interacting Hamiltonian $H$ pseudo-Hermitian as well. Equivalently, from Eqs. (\ref{eqn:POA}-\ref{eqn:CAFP}) and (\ref{eqn:EHE}) it is clear that such a term violates $\mathcal{C}\mathcal{P}\mathcal{T}$ symmetry.

Let us briefly consider this pseudo-Hermitian quartic interaction. A consequence of the pseudo-Hermiticity is that the usual Hermitian inner product $\langle \cdot | \cdot \rangle$ is no longer invariant under time evolution. Instead an invariant inner product is $\langle \cdot|\cdot\rangle_\eta \equiv \langle \cdot |\eta|\cdot \rangle$ or more generally $\langle \cdot |\eta\mathcal{O}|\cdot \rangle$, where $\mathcal{O}$ is a symmetry of the interacting theory (see e.g. Ref. \cite{Mostafazadeh:2002ph, Mostafazadeh:2003uq}). From Eq. (\ref{eqn:DEO}) one may check that $\langle \cdot|\cdot\rangle_\eta$ is indefinite, as states with odd numbers of spin down (anti)particles have negative square norm - they are ghosts - under this inner product. 

It has been shown that this pathology may be resolved, and a real spectrum is guaranteed, if there exists a symmetry $\mathcal{O}$ such that $\langle \cdot |\eta\mathcal{O}|\cdot \rangle$ is positive definite \cite{BenderBoettcher:1998rs, Mostafazadeh:2002ph}. However, by Eqs. (\ref{eqn:POA}) and (\ref{eqn:TCOA}), setting $\mathcal{O} = \mathcal{P}\mathcal{T}$ as in Ref. \cite{BenderBoettcher:1998rs} does not produce a positive definite inner product, and it appears to the authors that positive definiteness cannot be achieved by means of any other symmetry of the quartic interaction theory.

As is well-known, the properly normalized probability of the process $|\textrm{i}\rangle \to |\textrm{f}\rangle$ is
\begin{equation}
	P =  \frac{|\langle\textrm{f}|\mathcal{S}|\textrm{i}\rangle_\eta|^2}{\|\textrm{i}\|^2\|\textrm{f}\|^2}~,
\end{equation}
where $\|\cdot\|$ is the norm induced by $\langle \cdot |\cdot \rangle_\eta$ and $\mathcal{S}$ is the S-matrix. This probability must be negative whenever the number of spin down (anti)particles changes modulo 2 during the interaction, as then either $\|\textrm{i}\|^2$ or $\|\textrm{f}\|^2$ is negative, but not both. As such, the quartic interaction has ghost states that may participate in processes with negative probabilities. 

However, it is worth pointing out that in the non-relativistic low energy limit $|\bm{p}|^2 \ll m^2$, angular momentum conservation reduces to spin conservation. Further, in this limit there is no (anti)particle creation, so that the sum of particle and antiparticle numbers is conserved. Together with charge conservation, these conservation laws act as a superselection rule, such that they are sufficient to forbid processes with negative probabilities and the spin down states then become `physical ghosts' (see e.g. Ref. \cite{CrawfordBarut:1983nn}). The corresponding symmetry to this superselection rule must be $\eta$, since by Eq. (\ref{eqn:DEO}) it distinguishes between up and down spin states. It follows that in this limit $[\eta, H] = 0$, which implies that the interacting Hamiltonian is then Hermitian and the energy spectrum real. The quartic interaction is therefore an effective interacting theory in the non-relativisitic low energy limit: An appropriate limit in which to consider e.g. some condensed matter systems.

\subsection{Hermitian Theory}
Alternatively, let us consider the derivative coupling $\psit i\dsl\psi$. By the definition of the generalized adjoint field (\ref{eqn:DAF}) one may see that $\psit i\dsl \psi$ is manifestly Hermitian. Further from Eq. (\ref{eqn:FTFP}) we have
\begin{align}
	\langle0|T\big[i\slashed{\partial}_x\psi(x)\psit(y)\big]|0\rangle 
	& = i\slashed{\partial}_x\langle0|T \psi(x)\psit(y) |0\rangle\notag\\
	& = \int\!\!\!\frac{d^3p}{(2\pi)^3}\frac{i\slashed{p}}{p^2-m^2 + i\varepsilon}e^{-ip\cdot(x-y)}, \label{eqn:DIP}
\end{align}
where $\partial_x \equiv \partial/\partial x$, so that the time-ordered products of $\psit$ and $i\dsl \psi$ are well-defined in the Feynman prescription. Similarly $\{i\dsl\psi,\psit\} = 0$ outside the lightcone and the propagator is causal. We may therefore treat $\psit$ and $i\dsl\psi$ as canonical fields for the purpose of constructing an interacting theory.

A natural first choice for an interaction is the quartic term $(\psit i\slashed{\partial}\psi)^2$. This has mass dimension of four and hence by simple power counting we do not expect it to be renormalizable. Instead let us introduce a massive, real, scalar field $\phi$ and consider the derivative, Yukawa, Lagrangian
\begin{equation}
	\label{eqn:DCI}
	\mathcal{L}_{\textrm{int}} = g_1\psit (i\dsl\psi)\phi + g_2 \psit (i\dsl\psi)\phi^2.
\end{equation}
Based on na\"\i ve power counting, in $2+1$ dimensions we expect the first term to be renormalizable, the second term to be marginally relevant, and terms of this type with higher powers of $\phi$ must be non-renormalizable. For the remainder of this paper we briefly examine the properties of this derivative coupling theory.

\subsection{Symmetries}
Defining $\Psi$ as in Eq. (\ref{eqn:SSFT}) and applying the Fermi statistics, we may write
\begin{equation}
	\psit(i\dsl \psi) = \frac{g}{2}\Psi^T\begin{pmatrix} 0 & i\dsl \\ -i\overleftarrow{\dsl}^T & 0 \end{pmatrix} \Psi~,
\end{equation}
where $\overleftarrow{\dsl}$ means that the derivative acts to the left. The $\psit i \dsl\psi$ term therefore explicitly breaks the Sp(4) global symmetry of the free theory down to the Lorentz and fermion number symmetries: The global symmetry group is simply SO(2,1)$\times$U(1).

Applying Eqs. (\ref{eqn:POA}) and (\ref{eqn:TCOA}) along with their associated specific choice of $\gamma^\mu$ matrices, we further find that
\begin{align}
	\mathcal{P}_{i}\psit i\dsl\psi \mathcal{P}_{i} & = -\psit i\dsl\psi\notag\\
	\mathcal{T}\psit i\dsl\psi \mathcal{T} & = -\psit i\dsl\psi\notag\\
	\mathcal{C}\psit i\dsl\psi \mathcal{C} & = +\psit i\dsl\psi~,
\end{align}
so that the interacting theory violates parity and time-reversal, although $\mathcal{C}\mathcal{P}_i\mathcal{T}$ remains a symmetry as expected.

\subsection{Feynman rules and e$^-$e$^-$ scattering}
The Feynman rules for the interacting theory (\ref{eqn:DCI}) are generally trivial to obtain and similar to the usual $2+1$ Yukawa theory. One exception is the fermion propagator in Eq. (\ref{eqn:DIP}), and a second is that external fermion lines with spin index $s$ contracted with $i\dsl \psi$ obtain an extra factor $\pm(-1)^s m$, due to Eq. (\ref{eqn:SDE}). The extra sign plays a r\^ole in determining which spin combinations experience attractive interactions.

As an example, let us calculate amplitudes and cross-sections for electron-electron scattering at the tree level of Eq. (\ref{eqn:DCI}). (It is worth noting that the one loop diagrams $\sim g_2^2$ for this scattering are of the form $\int d^3p /p^4$ and so are also finite). The $t$-channel amplitude
\begin{equation}
i\mathcal{M}_t = \parbox[c][30mm]{30mm}{\includegraphics[scale=0.97]{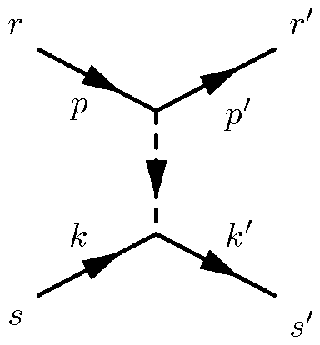}}~,
\end{equation}
where each external fermion line is labelled by its spin index and momentum. Applying the Feynman rules, we then have
\begin{equation}
i\mathcal{M}_t = -ig_1^2\frac{(-1)^{r+s}m^2}{t-M^2}\uhat^{s^\prime}(p^\prime)u^s(p)\uhat^{r^\prime}(k^\prime)u^r(k)~,
\end{equation}
where $M$ is the scalar mass. Assuming that the fermions are distinguishable for the sake of simplicity, then there is no $u$-channel contribution. In the non-relativistic limit $\uhat^{s^\prime}(p^\prime)u^s(p) = \delta^{s^\prime s}$, so that in the center of mass frame the amplitude becomes 
\begin{equation}
	\mathcal{M} = (-1)^{r+s}g_1^2m^2\frac{1}{|\bm{p}-\bm{p}^\prime|^2 + M^2}\delta^{s^\prime s}\delta^{r^\prime r}~.
\end{equation}
The extra $(-1)^{r+s}$ factor means that  spin up-up ($r=s=0$) and spin down-down ($r=s=1$) fermion interactions are attractive, while spin up-down ($r \not= s$) interactions are repulsive. Using the polarized completeness relations (\ref{eqn:SR2}) and the trace properties of the $\gamma^\mu$ matrices one may also calculate the (two-dimensional) scattering cross section.

\acknowledgments
This work is supported in part by the National Science Foundation.

\appendix
\section{Spinor Structure}
\label{app:SS}
In this appendix we derive the spinor structure of the free theory, as presented in the main body of the paper. To begin with, it is useful to note the following very well-known properties of the $2+1$ Dirac representation. First, the elements $\Lambda_{1/2}$ may be written as $\Lambda_{1/2} = \exp[-(i/2)\omega_{\mu\nu}S^{\mu\nu}]$ with the generators $S^{\mu\nu} = (i/4)[\gamma^\mu,\gamma^\nu]$. The $\gamma^{\mu}$ matrices are the $2\times2$ traceless (anti)Hermitian matrices with appropriate factors of $i$ such that $\{\gamma^\mu, \gamma^\nu\} = 2\eta^{\mu\nu}\times\id$, and the fermionic field $\psi$ therefore has two components. As usual $\big(\gamma^0\big)^2 = \id$, $\big(\gamma^i\big)^2 = -\id$ and $\gamma^{0\dagger} = \gamma^0$, $\gamma^{i\dagger} = -\gamma^{i}$, or equivalently $\gamma^{\mu\dagger} = \gamma^0\gamma^\mu\gamma^0$. (E.g. a possible choice is $\gamma^0 = \sigma_z$, $\gamma^1 = i\sigma_x$ and $\gamma^2 = i\sigma_y$, where $\sigma_{x,y,z}$ are the usual $2\times 2$ Pauli matrices, but we do not make a specific choice in this appendix.) Finally, the $\gamma^{\mu}$ matrices have the well-known commutation and Lorentz transformation properties (see e.g. Refs. \cite{Weinberg:1995qf,PeskinSchroeder:1995qf})
\begin{align}
	\Lambda_{\frac{1}{2}}^{\dagger}\gamma^0 & = \gamma^0\Lambda_{\frac{1}{2}}^{-1}~,\label{eqn:GZCR}\\
	\Lambda: \gamma^{\mu} & \mapsto \Lambda_{\frac{1}{2}}\big(\Lambda^{\mu}_{~\nu}\gamma^{\nu}\big)\Lambda_{\frac{1}{2}}^{-1} = \gamma^{\mu}~,\label{eqn:GMTL}
\end{align}
the latter of course implying $\gamma^{\mu}$ are invariants under Lorentz transformations.

To derive the spinor structure, we need only consider the classical field theory corresponding to $\mathcal{L}_0$. The general solution of Eq. (\ref{eqn:FTEM}) must clearly be of the form $\psi(x) = u(p)e^{-ip\cdot x} + v(p)e^{+ip\cdot x}$ with dispersion relation $p^2 = m^2$, and we require $u(p), v(p) \in \mathbb{C}^2$ to be spinors which transform according to Eq. (\ref{eqn:TFF}). That is we require $u(\Lambda p) = \Lambda_{1/2} u(p)$. In the case of the Dirac theory one considers the positive and negative energy solutions separately, because the Dirac equation of motion acting on the fields provides the extra constraint $(\slashed{p} - m)u(p) =  (\slashed{p} + m)v(p) = 0$, which fixes the positive and negative energy spinors, $u(p)$ and $v(p)$, to distinct, orthogonal (with respect to the Dirac inner product) subspaces of $\mathbb{C}^2$.  However, for our present theory there is no Dirac equation constraint on the fields, so there is no motivation to consider positive and negative energy solutions independently. Instead, all we may say is that $u(p)$ and $v(p)$ must both belong to the same inner product (Hilbert) space on $\mathbb{C}^2$. This observation is key to obtaining the inner products and completeness relations presented in Sec. \ref{sec:LFS}. We are therefore free to choose the same spinor basis for both $u(p)$ and $v(p)$, so we may then write the general solution to the equation of motion as
\begin{equation}
	\label{eqn:GSEMFT}
	\psi_{\cl}(x) = \!\!\!\int \!\!\!\frac{d^2 \bm{p} }{(2\pi)^2}\frac{1}{\sqrt{2E_{\bm{p}}}} \sum_{s=0,1}u^s(p)\Big(\alpha^s_{\bm{p}}e^{-ip\cdot x} + \beta^{s\dagger}_{\bm{p}}e^{+ip\cdot x}\Big)~,
\end{equation}
where $\{u^s(p)\}$ is the abovementioned spinor basis of $\mathbb{C}^2$, $\alpha^s_{\bm{p}}$ and $\beta^{s\dagger}_{\bm{p}}$ are Fourier coefficients, and the subscript `cl' denotes a classical field. In accordance with Eq. (\ref{eqn:GSEMFT}), an ansatz for the adjoint field is
\begin{equation}
	\label{eqn:GSEMFT2}
	\psit_{\cl}(x) = \!\!\!\int \!\!\!\frac{d^2 \bm{p} }{(2\pi)^2}\frac{1}{\sqrt{2E_{\bm{p}}}}\! \sum_{s=0,1}\uhat^s(p)\Big(\alpha^{s\dagger}_{\bm{p}}e^{ip\cdot x} + \beta^{s}_{\bm{p}}e^{-ip\cdot x}\Big)~,
\end{equation}
in which we have anticipated that the adjoint will involve an Hermitian conjugation together with some linear operation on the spin indicies. We call $\uhat^s(p)$ the adjoint spinor, an expression for which we now seek to find.

A judicious choice of the spinor basis renders the analysis of the spinors much simpler. A convenient choice is to choose the basis spinors $u^s(p)$ to be the solutions of the positive and negative energy Dirac spinor equations, $(-1)^s\slashed{p}u^s = mu^s$. Note that these equations do not arise out of an equation of motion: we have arbitrarily introduced them since their solutions span $\mathbb{C}^2$ and hence form a basis. To prove that we may always make such a choice even without a Dirac equation of motion, consider the $\gamma^0$ eigenvectors $\xi^s$, $s=0,1$, having unit norm $\xi^{s\dagger}\xi^r = \delta^{rs}$. Under the normalization chosen for $\gamma^0$, we have
\begin{equation}
	\label{eqn:DGZE}
	\gamma^0\xi^s = (-1)^{s}\xi^s~.
\end{equation}
Letting $p_0 = (m,\bm{0})$ be the rest frame momentum and employing the usual Feynman slash notation, then Eq. (\ref{eqn:DGZE}) is equivalent to $(\slashed{p}_0/m)\xi^s = (-1)^{s}\xi^s$, so clearly Eq. (\ref{eqn:DGZE}) is nothing but the rest frame Dirac equation. Boosting both sides, since $\gamma^\mu$ are Lorentz invariants (\ref{eqn:GMTL}), one finds
\begin{equation}
	\label{eqn:SDE2}
	\big(\slashed{p}/m\big)\Lambda_{\frac{1}{2}}\xi^s =  (-1)^{s}\Lambda_{\frac{1}{2}}\xi^s~.
\end{equation}
where $p = \Lambda p_0$. Let us henceforth choose the spinor basis elements $u^s(p)$ arising in Eq. (\ref{eqn:GSEMFT}) to be the boosts of $\xi^s$. That is, let us define
\begin{equation}
	\label{eqn:DTL}
	u^s(p) \equiv \Lambda_{\frac{1}{2}}\xi^s~.
\end{equation}
Then from Eq. (\ref{eqn:SDE2}) it follows that under this choice
\begin{equation}
	\label{eqn:SDE3}
	\big(\slashed{p}/m\big)u^s(p) = (-1)^su^s(p)~,
\end{equation}
which is Eq. (\ref{eqn:SDE}). Hence the spinor basis choice (\ref{eqn:DTL}) are the solutions to the Dirac equation as claimed. It is important to note once more that although Eq. (\ref{eqn:SDE3}) is the Dirac equation, it holds here only for our particular choice of basis spinors (\ref{eqn:DTL}). It does not hold for a general spinor in the solution to the equation of motion (\ref{eqn:FTEM}). The advantage of this choice is that not only are the properties of $u^s(p)$ solely determined by the properties of $\Lambda_{1/2}$ and $\xi^s$, but also that Eq. (\ref{eqn:SDE}) or (\ref{eqn:SDE3}) will prove to be a useful algebraic relation. Note also that implicit in Eq. (\ref{eqn:DTL}) is the transformation law
\begin{equation}
	u^s(\Lambda p) = \Lambda_{\frac{1}{2}}u^s(p)~.
\end{equation}

Suppose the adjoint $\uhat^s(p)$ is chosen to be the Dirac adjoint, i.e.  $\uhat^s(p) = \bar{u}^s(p) \equiv u^{s\dagger}(p)\gamma^0$. Applying Eqs. (\ref{eqn:GZCR}), (\ref{eqn:DGZE}) and (\ref{eqn:DTL}) one finds that the corresponding Lorentz invariant inner product of basis spinors - the normalization relation - is
\begin{equation}
	\label{eqn:DLIP}
	\bar{u}^r(p)u^s(p) = \xi^{r\dagger}\gamma^0\xi^s = (-1)^s\delta^{rs}~,
\end{equation}
which is clearly indefinite. But, the Lorentz invariant inner product must be positive (or negative) definite in order for consistent spinor completeness relations to exist. 

To remedy this problem, we note that we may choose the adjoint spinor to have a more general form
\begin{equation}
	\label{eqn:DGAS}
	\uhat(p) = \Big[\Gammap u(p)\Big]^\dagger\gamma^0
\end{equation}
where $\Gammap$ is an an invertible linear operator $\Gammap:\mathbb{C}^2 \to \mathbb{C}^2$ - a  tensor on the spin indices - which transforms under the Lorentz transformation as 
\begin{equation}
	\label{eqn:TLLM}
	\Gammap \mapsto \Gamma_{\Lambda p} = \Lambda_{\frac{1}{2}}\Gammap\Lambda_{\frac{1}{2}}^{-1}~.
\end{equation}
Such a choice of the adjoint spinor is permitted because under a Lorentz transformation, we have 
\begin{equation}
	\label{eqn:ASTL}
	\uhat^s(p) \mapsto \uhat^s(\Lambda p) = u^s(p)^\dagger\Lambda_{\frac{1}{2}}^\dagger\Lambda_{\frac{1}{2}}^{-\dagger}\Gammap^\dagger\Lambda_{\frac{1}{2}}^\dagger\gamma^0 = \uhat^s(p)\Lambda_{\frac{1}{2}}^{-1}~, 
\end{equation}
as required by Eq. (\ref{eqn:TFF}). Hence we have Eq. (\ref{eqn:SLT}).

Now, due to the contraction over the Lorentz vector indices, $\slashed{p}$ is by definition a tensor transforming according to Eq. (\ref{eqn:TLLM}). Choosing
\begin{equation}
	\label{eqn:CGM}
	\Gammap = \slashed{p}/m~,
\end{equation}
the relation $\uhat^s(p) = (-1)^s\bar{u}(p)$, which is Eq. (\ref{eqn:SR1}), is obtained from the Dirac equation (\ref{eqn:SDE3}). Applying (\ref{eqn:DLIP}) it then follows that the Lorentz inner product becomes simply $\uhat^r(p)u^s(p) = \delta^{rs}$, which is positive definite and is Eq. (\ref{eqn:SNR}). Further, by Eqs. (\ref{eqn:DTL}), (\ref{eqn:DGAS}) and (\ref{eqn:ASTL}) the corresponding completeness relations for this inner product are
\begin{equation}
	\sum_{s= 1,2}u_a^s(p)\uhat_b^s(p) = \bigg(\sum_{s= 1,2}\Lambda_{\frac{1}{2}}\xi^s\widetilde{\xi}^{s}\Lambda_{\frac{1}{2}}^{-1}\bigg)_{ab}  = \delta_{ab}~,
\end{equation}
as claimed in Eq. (\ref{eqn:SCR}). Note that even though the rest frame adjoint is now $\widetilde{\xi}^{s} = \xi^{s\dagger}(\slashed{p}_0^\dagger/m)\gamma^0 = \xi^{s\dagger}$, $\widetilde{\xi}^s$ does not transform as $\xi^{s\dagger}$ under $\Lambda$, because $\slashed{p}_0$ remains a tensor under $\Lambda$, whereas $\gamma^0$ is a Lorentz invariant.

From Eqs. (\ref{eqn:SDE} - \ref{eqn:SCR}) follow the useful relations
\begin{align}
	\big[\uhat^r(p)u^s(k)\big]^\dagger & = (-1)^{r+s}\uhat^s(k)u^r(p)~,\label{eqn:SR3}\\
	u^r(p)\uhat^r(p) & = \big[\id + (-1)^r\slashed{p}/m\big]/2~,\label{eqn:SR2}
\end{align}
the latter of which is derived by noting $\sum_s(-1)^su^s\uhat^s = \slashed{p}/m\sum_su^s\uhat^s$. The former implies that the inner product $\uhat u$ is not well-defined with respect to Hermitian conjugation. These relations will prove useful e.g. in the calculation of interaction amplitudes. It is also useful to note that
\begin{align}
	u^{s\dagger}(p)u^s(p)
	& = \xi^{s\dagger}\Lambda_{\frac{1}{2}}^\dagger\Lambda_{\frac{1}{2}}\xi^s \notag\\
	& = \xi^{s\dagger}\big[\Lambda_{\frac{1}{2}}^\dagger\Lambda_{\frac{1}{2}}\big]^{-1}\xi^s~,
\end{align}
where the second line is obtained by inserting an $\id = (\gamma^0)^2$ and applying Eqs. (\ref{eqn:GZCR}) and (\ref{eqn:DGZE}). This relation together with the positive definiteness of $\Lambda_{1/2}^\dagger\Lambda_{1/2}$ and Eq. (\ref{eqn:SR1}) is sufficient to imply
\begin{equation}
	\label{eqn:SR4}
	\uhat^0(p)\gamma^0u^0(p) = -\uhat^1(p)\gamma^0u^1(p)~,
\end{equation}
which is needed e.g. to show $J|0\rangle = 0$ from Eq. (\ref{eqn:FEAO}).

\end{document}